\documentclass[preprint,showpacs,preprintnumbers,amsmath,amssymb]{revtex4}

\usepackage{graphicx}
\usepackage{dcolumn}
\usepackage{bm}

\newcommand{\partialF}{\mbox{$\not{\!\partial}$}}

\newcommand{\braketo}[3]{\mbox{$\langle #1 | #2 | #3 \rangle$}}
\newcommand{\braketi}[2]{\mbox{$\langle #1 | #2 \rangle$}}

\newcommand{\ket}[1]{\mbox{$| #1 \rangle$}}
\newcommand{\comm}[2]{\mbox{$[#1,#2]$}}
\newcommand{\abs}[1]{\mbox{$\left| #1 \right|$}}

\begin{document}

\preprint{}

\title{Instability of the hedgehog shape for the octet baryon
in the chiral quark soliton model}

\author{Satoru Akiyama}
\email{akiyama@ph.noda.tus.ac.jp}
\author{Yasuhiko Futami}
\affiliation{
	Department of Physics, Tokyo University of Science,
	2641, Noda, Chiba 278-8510, Japan 
}

\date{\today}

\begin{abstract}
In this paper the stability of the hedgehog shape of the chiral soliton is studied 
for the octet baryon with the SU(3) chiral quark soliton model.
The strangeness degrees of freedom are treated by a simplified bound-state approach,
which omits the locality of the kaon wave function.
The mean field approximation for the flavor rotation is applied to the model.
The classical soliton changes shape according to the strangeness.
The baryon appears as a rotational band of the combined system
of the deformed soliton and the kaon.
\end{abstract}


\maketitle

\section{Introduction\label{sec:intro}}
In the limit of a large number of colours $N_{c}$ of QCD  \cite{rf:tHooft74}
the baryon appears as a soliton \cite{rf:Witten79,rf:Dashen94}.
By now, various effective models of QCD in the low energy region
have employed the soliton picture of the baryon:
the Skyrme model \cite{rf:Skyrme61,rf:Adkin83,rf:Guadanini84,rf:Mazur84,rf:Yabu88,rf:Balachandran85},
the Nambu-Jona-Lasinio (NJL) model \cite{rf:Nambu61,rf:Ebert86,rf:Weigel92}
and the chiral quark soliton model (CQSM)
 \cite{rf:Diakonov88,rf:Reinhardt88,rf:Reinhardt89,rf:Wakamatsu91}.
The static properties (mass, magnetic moment) \cite{rf:Chistov96,rf:Alkofer96,rf:Wakamatsu96}
and the quark distribution functions \cite{rf:Diakonov96,rf:Weigel97,rf:Wakamatsu98}
of the baryon have been studied by means of these models.
Here, the baryon appears as a rotational band of the soliton \cite{rf:Adkin83}.
The soliton takes the hedgehog shape and adiabatically rotates
as a rigid body in flavor and real space.
Although there is also an approach using the harmonic approximation of the meson fluctuations
off the shape \cite{rf:Weigel93}, in any case the stability of the shape is assumed.

From the point of view of the constituent quark, the assumption is justified
in the case of the nucleon ($N$), since the hedgehog shape of the chiral meson fields
is caused by the S state of the u,d quarks
which have equal mass \cite{rf:Chodos75}.
However it may be not so for the strange baryons ($\Lambda$, $\Sigma$, $\Xi$, etc.)
consisting of quarks with different masses.  
The inertial force in the body fixed frame of the soliton depends on the mass.
Therefore, the hedgehog shape would be unstable.

The baryon as a nonrigid rotator has been studied with the Skyrmion \cite{rf:Hosaka01}.
There, the effects of the Coriolis force were neglected in the shape of the static soliton
and included perturbatively in the state vector for the collective rotation. 
For the excited states of the baryon, there are several studies of this problem
in the nonrelativistic quark model \cite{rf:Takayama99}.

In this article we shall argue this problem for the octet baryon by means of the CQSM.
This model provides the simplest foothold to study the above problem from the point of view
of the relativistic quark model.
The shape of the soliton can be determined self-consistently as the mean field potential
for the quark. Thus the model can eliminate the influence of the artificial assumption 
that the mean field potential takes a spherical shape.

In Sec.~\ref{sec:model}, after a brief review of the SU(3) chiral quark soliton model
we show the cranking method for the deformed soliton
and the mean field approximation for the rotated system.
Section~\ref{sec:perturb} deals with the perturbative expansion of the effective action.
At the the lowest order expansion, we show the profile functions 
and the energy of the deformed classical soliton.
For the higher order expansion, we construct the Hamiltonian due to 
the collective motion of the soliton and the canonical quntization of the collective variables.
In Sec.~\ref{sec:diagonal}, first we deal with the kaon in the background soliton and
the stability of the combined system of the soliton and the kaon.
Next we give an outline of the diagonalization of the collective Hamiltonian
and show the results for the octet baryon.
Finally, in Sec.~\ref{sec:discuss} we summarize the results and
discuss the relation to other papers.

\section{The SU(3) chiral quark soliton model and its mean field approximation \label{sec:model}}
The chiral quark model in case of the flavor SU(3) is given
by the following path integral \cite{rf:Diakonov88,rf:Reinhardt88,rf:Blotz93}.
\begin{equation}
	Z = \int D\psi\,D\bar{\psi}\,DU 
		\exp \left[
			i \int d^{4}x \bar{\psi} \left(
				i\partialF-MU^{\gamma_{5}}-\hat{m}
			\right) \psi
		\right]
\end{equation}
with
\begin{equation}
	U^{\gamma_{5}}(x) = 
		\frac{1+\gamma_{5}}{2} U(x)+ 
		\frac{1-\gamma_{5}}{2} U^{\dagger}(x),
\end{equation}
where $\psi$ is the quark field
and $U$ is the chiral meson field $\in$ SU(3).
Furthermore, $M$ is the dynamical quark mass
and $\hat{m}$ is the current quark mass matrix given by
\begin{eqnarray}
	\hat{m} &=& m_{0} \lambda_{0} + m_{3} \lambda_{3} + m_{8} \lambda_{8}\nonumber\\
		&=& \left(
			\begin{array}{ccc}
				m_{u} &     0 &     0\\
			 	    0 & m_{d} &     0\\
			 	    0 &     0 & m_{s}\\
			\end{array}
		\right),
\end{eqnarray}
where $\lambda_{\mu}\,(\mu = 1,2,\dots,8)$ are the Gell-Mann matrices,
$\lambda_{0}=(\sqrt{2/3})\,{\bf 1}$,
and $m_{3} \approx 0$ because $m_{u} \approx m_{d} \ll m_{s}$.
In this article, we set $M = 400$~MeV, $m_{u} = m_{d} = 6$~MeV,
and $m_{s} = 200$~MeV. 

Using the path integral formula for the quark field, we find
\begin{eqnarray}
	Z &=& \int DU e^{i S_{F}\left[ U \right]},\\
	i S_{F}\left[ U \right] &=& N_{c} \log\det \left(i\partial_{t}-H\right),
\label{eq:logdet}
\end{eqnarray}
where $N_{c}$ is the number of colors and
\begin{equation}
	H = \frac{1}{i} {\bm \alpha} \cdot \nabla + 
		\beta \left(MU^{\gamma_{5}}+\hat{m}\right)
\end{equation}
is the quark Hamiltonian.

For the chiral field $U(x)$, we postulate the so called cranking form
 \cite{rf:Adkin83,rf:Braaten88}:
\begin{equation}
	U(x) = {\cal A}(t)B^{\dagger}(t) U_0({\bf r}) B(t){\cal A}^{\dagger}(t),
\label{eq:aua}
\end{equation}
where $U_0({\bf r})$ is the static chiral meson field,
${\cal A}(t)$ describes the adiabatic rotation of the system in SU(3) flavor space,
and $B(t)$ describes the spatial rotation. 
We write $U_0({\bf r})$ as
\begin{equation}
	U_0({\bf r})
	= \left(
		\begin{array}{cc}
			e^{i F({\bf r})\, \hat{\bf \Lambda}({\bf r}) \cdot {\bm \tau}} & 0\\
			0^{\dagger} & 1
		\end{array}
	\right),
\label{eq:u0_embed}
\end{equation}
where $F$ is the radial component of the profile function,
$\hat{\bf \Lambda}$ is a unit vector in the isospin space,
and ${\bm \tau}$ represents Pauli matrices.
For the flavor rotation we write \cite{rf:Kaplan90}
\begin{equation}
	{\cal A}(t) = \left(
				\begin{array}{cc}
					A(t)		& 0\\
					0^{\dagger} 	& 1
				\end{array}
			\right) A_{s}(t),
\label{eq:A_param}
\end{equation}
where $A$ is the flavor SU(2) rotation operator and 
$A_{s}$ rotates $U_{0}$ into the strange directions.
In particular, we parametrize $A_{s}(t)$ as
\begin{equation}
	A_{s}(t) = \exp i {\cal D}(t),
\label{eq:S_param}
\end{equation}
where
\begin{equation}
	{\cal D}(t) = \left(
				\begin{array}{cc}
				0 & \sqrt{2} D(t)\\
				\sqrt{2} D^{\dagger}(t) & 0
				\end{array}
			\right)
\label{eq:isodublet_D}
\end{equation}
and $D = (D^{1}, D^{2})^{T}$ is the isodoublet spinor. 
Although Eq.~(\ref{eq:A_param}) was motivated by the bound-state approach \cite{rf:Callan85},
we will not treat the locality of the kaon wave function in this article.

From Eqs.~(\ref{eq:logdet}) and (\ref{eq:aua}),
we obtain
\begin{equation}
	i S_{F} = N_{c} 
		\log\det \left(
			i\partial_{t}
			+i {\cal A}^{\dagger}\dot{\cal A}
			+i B\dot{B}^{\dagger}
			-H'
		\right),
\label{eq:rot_s}
\end{equation}
where
\begin{equation}
	H' = {\cal A}^{\dagger} B H B^{\dagger} {\cal A}
 		= \frac{1}{i} {\bm \alpha} \cdot \nabla + 
		\beta \left(MU^{\gamma_{5}}_{0}+\hat{m}'\right) 
\label{eq:rot_h}
\end{equation}
is the rotated Hamiltonian and $\hat{m}'$ is given by
\begin{eqnarray}
	\hat{m}' &=& {\cal A}^{\dagger}\hat{m}{\cal A}
		= m_{0} \lambda_{0} + m_{8} D^{(8)}_{8\mu}(A_{s}) \lambda_{\mu}.
	\label{eq:rot_mq}
\end{eqnarray}
Here, the term proportional to $m_{8}$ breaks the SU(3) symmetry and
$D^{(8)}_{8\mu}(A_{s})$ is the Wigner \textit{D} function of $A_{s}$ in the adjoint representation.
We show some important components,
\begin{eqnarray}
	D^{(8)}_{83} &=& - \frac{\sqrt{3}}{2}
		\frac{\sin^2 \sqrt{\kappa_{0}}}{\kappa_{0}} \kappa_{3},\\
	D^{(8)}_{88} &=& 1 - \frac{3}{2} \sin^2 \sqrt{\kappa_{0}},
\end{eqnarray}
where
\begin{eqnarray}
	\kappa_{0} &=& 2D^{\dagger} D,\\
	\kappa_{3} &=& 2D^{\dagger} \tau_{3} D.
\end{eqnarray}

\begin{figure}
	\begin{center}
		\includegraphics{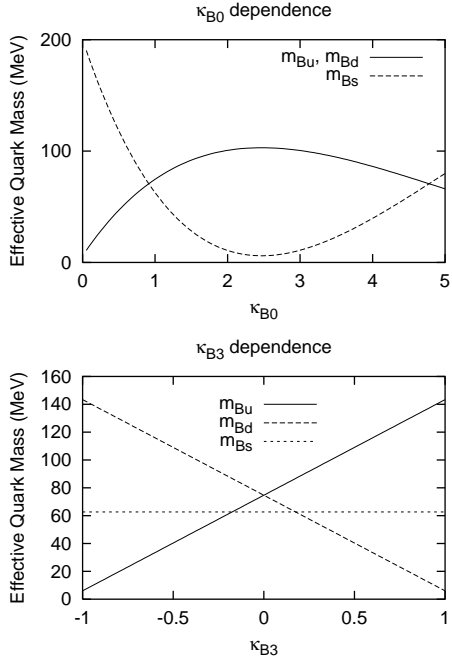}\\
	\end{center}
	\caption{The $\kappa_{B0}$ and $\kappa_{B3}$ dependencies of
	the effective quark mass.
	For $\kappa_{B0}$ dependence, we set $\kappa_{B3} = 0$,
	at which $m_{Bu}$ and $m_{Bd}$ take the same value.
	For $\kappa_{B3}$, we set $\kappa_{B0} = 1$.}
\label{fig:meff}
\end{figure}

We now wish to discuss this model in the mean field approximation.
Suppose that the collective motion $\cal A$ and $B$ is quantized
and $\ket{B}$ as an eigenstate of the collective Hamiltonian.
The operators $D$ and $D^{\dagger}$ convert a non-strange quark into a strange one.
Therefore, if \ket{B} points to a specific direction in the isospin space,
\begin{eqnarray}
 	\kappa_{B0} &=& \braketo{B}{\kappa_{0}}{B},
\label{eq:ddT0}\\
	\kappa_{B3} &=& \braketo{B}{\kappa_{3}}{B}
\label{eq:ddT3}
\end{eqnarray}
have nonzero values. 
For a fixed $\kappa_{B0}$, we have the following constraint for $\kappa_{B3}$, 
\begin{equation}
	- \kappa_{B0} \leq \kappa_{B3} \leq + \kappa_{B0}.
\label{eq:ddTrange}
\end{equation}
Intuitively, $\kappa_{B0}$ and $\kappa_{B3}$ represent the quantity of the strangeness,
respectively, and the asymmetry of the strangeness coupled to the isodoublet in $\ket{B}$.
The expectation value $\braketo{B}{\hat{m}'}{B}$ may be approximated by
\begin{eqnarray}
	\hat{m}_{B} &=& m_{0} \lambda_{0}
		+ m_{B3} \lambda_{3} + m_{B8} \lambda_{8} \nonumber\\
	&=& \left(
		\begin{array}{ccc}
			m_{Bu} & 0 & 0\\
		 	0 & m_{Bd} & 0\\
		 	0 & 0 & m_{Bs}\\
		\end{array}
	\right)
\label{eq:mq_eff}
\end{eqnarray}
with
\begin{equation}
	m_{B\mu} = 
	m_{8} \lim_{\kappa_{0,3} \rightarrow \kappa_{B0,3}}{D^{(8)}_{8\mu}(A_{s})}
	\,\,\,(\mu = 3,8).\\ 
\end{equation}
Equation~(\ref{eq:mq_eff}) indicates that the u,d,s quarks mix with each other
by the rotation in the strange direction and their masses are renormalized
in the body fixed frame.
Figure~\ref{fig:meff} shows the $\kappa_{B0}$ and $\kappa_{B3}$ dependencies of 
the effective quark masses $m_{Bu,d,s}$.

\section{Perturbation and canonical quantization\label{sec:perturb}}
We separate $H'$ into an unperturbed part $H'_{0}$ and a perturbing part $\Delta H'$ as follows:
\begin{eqnarray}
	H' &=& H'_{0} + \Delta H',
\label{eq:htot}\\
	H'_{0} &=& \frac{1}{i} {\bm \alpha} \cdot \nabla + 
		\beta \left(MU^{\gamma_{5}}+\hat{m}_{B}\right),
\label{eq:h0}\\
	\Delta H' &=& \beta \left(\hat{m}' - \hat{m}_{B}\right).
\label{eq:h1}
\end{eqnarray}
We expand $iS_{F}$ in powers of $i {\cal A}^{\dagger}\dot{\cal A}$,
$i B\dot{B}^{\dagger}$, and $\Delta H'$
around the eigenstate of $(i\partial_{t}-H'_{0})$ \cite{rf:Reinhardt89}.

Since $H'_{0}$ contains the $\hat{m}_{B}$, $H'_{0}$ breaks not only the SU(3) symmetry
but also SU(2). Thus it has only rotation symmetry around the third axis in isospin space.
When the isospin space is mapped into the real space,
the configuration of the system has an axial symmetry with respect 
to the specific axis in real space.

The grand spin operator in the quark space is given by
\begin{equation}
	{\bf K}^{(q)} = {\bf J}^{(q)} + {\bf I}^{(q)}
			= {\bf L}^{(q)} + {\bf S}^{(q)} + {\bf I}^{(q)},
\end{equation}
where ${\bf J}^{(q)}$, ${\bf L}^{(q)}$, and ${\bf S}^{(q)}$ are
the total, orbital, and spin angular momenta, respectively,
and ${\bf I}^{(q)}$ denotes the isospin.
The above argument is expressed as
\begin{equation}
	\comm{K^{(q)}_{3}}{U_{0}({\bf r})} = 0,
\label{eq:k3_symmetry}
\end{equation}
where $K^{(q)}_{3}$ is the third component of ${\bf K}^{(q)}$ and 
we choose the $z$ axis as the specific one. 
We call this equation ``$K_{3}$ symmetry''.

The isosinglet part of Eq.~(\ref{eq:k3_symmetry}) means that
the radial component of the profile function $F$ depends only on 
$r$ and $\theta$ of the polar coordinates.
On the other hand, the isotriplet part means that the unit vector $\hat{\bf \Lambda}$
has axial symmetry about $z$ axis.
Thus, the deviation from the hedgehog shape takes place
only in the $r$ and $\theta$ directions.

In actual calculation, we introduce the tensor operators $Z^{JLSI}_{KK_{3}}({\bf \hat{r}})$
defined by the following equations, where $r = \left| {\bf r} \right|$ and
${\bf \hat{r}} = {\bf r}/r$ \cite{rf:Oh94}.
\begin{eqnarray}
	&&\comm{K^{(q)}_{3}}{Z^{JLSI}_{KK_{3}}} = K_{3} Z^{JLSI}_{KK_{3}},\\
	&&\comm{{\bf K}^{(q)2}}{Z^{JLSI}_{KK_{3}}} = K(K+1) Z^{JLSI}_{KK_{3}},\\
	&&\comm{{\bf J}^{(q)2}}{Z^{JLSI}_{KK_{3}}} = J(J+1) Z^{JLSI}_{KK_{3}},\\
	&&\comm{{\bf L}^{(q)2}}{Z^{JLSI}_{KK_{3}}} = L(L+1) Z^{JLSI}_{KK_{3}},\\
	&&\comm{{\bf S}^{(q)2}}{Z^{JLSI}_{KK_{3}}} = S(S+1) Z^{JLSI}_{KK_{3}},\\
	&&\comm{{\bf I}^{(q)2}}{Z^{JLSI}_{KK_{3}}} = I(I+1) Z^{JLSI}_{KK_{3}}.
\end{eqnarray}
Then the chiral fields can be expanded in the series of $Z^{JLSI}_{KK_{3}}$:
\begin{equation}
	U_{0}({\bf r})
	= \sum_{L} \sum_{I} \sum_{K} \sum_{K_{3}} 
		U^{LL0I}_{KK_{3}}(r) Z^{LL0I}_{KK_{3}}({\bf \hat{r}}).
\label{eq:mode_U}\\
\end{equation}
Equation~(\ref{eq:k3_symmetry}) means that Eq.~(\ref{eq:mode_U})
contain only $K_{3} = 0$ components.

Similarly, the quark fields are expanded in the Kana-Ripka bases
$\phi_{KK_{3}\alpha}$ \cite{Kahana84}:
\begin{equation}
	\psi({\bf r}) = \sum_{KK_{3}\alpha} C_{KK_{3}\alpha} \phi_{KK_{3}\alpha}({\bf r}),
\end{equation}
where the subscript $\alpha$ indicates quantum numbers other than $K$ and $K_{3}$.
The matrix elements of $H'_{0}$ with $\phi_{KK_{3}\alpha}$
have a contribution only from the bases with the same $K_{3}$.

\subsection{The self-consistent classical soliton\label{sec:s0}}
The lowest order perturbative expansion of
the action $iS_{F}$ is given by
\begin{equation}
	i S_{F0} = N_{c} \log\det \left(i\partial_{t}-H'_{0}\right) 
		= - iE_{cl}T,
\label{eq:rot_s0}
\end{equation}
where $T$ is a sufficiently large time interval
and $E_{cl}$ is the classical soliton energy.
With the eigenvalues $\epsilon_{n}$ of $H'_{0}$ and
$\epsilon^{0}_{n}$ of
\begin{equation}
	K' = \lim_{U_{0} \rightarrow 1} H'_{0} 
	= \frac{1}{i} {\bm \alpha} \cdot \nabla + 
		\beta \left(M + \hat{m}_{B}\right),
\label{eq:kinetic}
\end{equation}
we obtain
\begin{equation}
	E_{cl}	= \sum_{n} \left[N_{c} \eta^{val}_{n} + \rho^{vac}_{\Lambda}(\epsilon_{n})\right] \epsilon_{n}
			- \sum_{n} \rho^{vac}_{\Lambda}(\epsilon^{0}_{n}) \epsilon^{0}_{n},
\end{equation}
where $\eta^{val}_{n}$ is the occupation number of the valence quark level $\epsilon_{n}$
and $\rho^{vac}_{\Lambda}(\epsilon_{n})$ is the cutoff function of the vacuum energy
for level $\epsilon_{n}$ with cutoff parameter $\Lambda$ (Appendix).
In this article, we use Schwinger's proper time regularization \cite{rf:Ebert86,rf:Diakonov88,rf:Schwinger51}
and set $\Lambda = 700$~MeV,
and we assume that the valence quark is in the lowest positive energy state.

To subtract $E_{cl}$, we use $K'$ but
\begin{equation}
	K = \lim_{U \rightarrow 1} H =
		 \frac{1}{i} {\bm \alpha} \cdot \nabla + \beta \left(M + \hat{m}\right).
\end{equation}
The reason is the following.
Physically, $E_{cl}$ is the static soliton energy in the body fixed frame and
its subtraction point is the Dirac sea in the absence of a static soliton.
We employ the body fixed frame in which $\kappa_{B0}$ and $\kappa_{B3}$ are nonzero.
In this frame the Hamiltonian without the static soliton is $K'$.

The equations of motion for the profile functions \cite{rf:Reinhardt88}
are obtained by the extremum conditions
for the action $S_{F0}$ with respect to the radial component $F$
and the direction of the unit vector $\hat{\bf \Lambda}$:
\begin{eqnarray}
	&&S({\bf r}) \sin F({\bf r}) = {\bf P}({\bf r}) \cdot \hat{\bf \Lambda}({\bf r}) \cos F({\bf r}),
\label{eq:eom_f}\\
	&&\hat{\bf \Lambda}({\bf r}) = {\rm sgn} \left[S({\bf r}) \sin 2 F({\bf r})\right]
				\frac{{\bf P}({\bf r})}{\abs{{\bf P}({\bf r})}}.
\label{eq:eom_angle}
\end{eqnarray}
Here, $S({\bf r})$ and ${\bf P}({\bf r})$ are  the scalar-isoscalar
and pseudoscalar-isovector densities, respectively, and are defined by
\begin{eqnarray}
	S({\bf r}) &=& \sum_{n} \rho_{\Lambda}^{R}(\epsilon_{n})\,
		\bar{\psi}_{n}({\bf r}) \psi_{n}({\bf r}),\\
	{\bf P}({\bf r}) &=& \sum_{n} \rho_{\Lambda}^{R}(\epsilon_{n})\,
		\bar{\psi}_{n}({\bf r}) i\gamma_{5} {\bm \tau} \psi_{n}({\bf r}),
\end{eqnarray}
where $\psi_{n}({\bf r}) = \braketi{{\bf r}}{n}$, $\ket{n}$ is the eigenvector
belonging to the eigenvalue $\epsilon_{n}$ of $H'_{0}$,
and $\rho_{\Lambda}^{R}(\epsilon_{n})$ is
the cutoff function shown in the Appendix.
Using the boundary conditions
\begin{eqnarray}
	\lim_{r \rightarrow 0} F({\bf r}) &=& -\pi,\\
	\lim_{r \rightarrow \infty} F({\bf r}) &=& 0,
\end{eqnarray}
Eqs.~(\ref{eq:eom_f}),(\ref{eq:eom_angle}) and the Dirac equation with
the Hamiltonian Eq.~(\ref{eq:h0}) are self-consistently solved \cite{rf:Reinhardt88}.

\begin{figure}
	\begin{center}
		\includegraphics{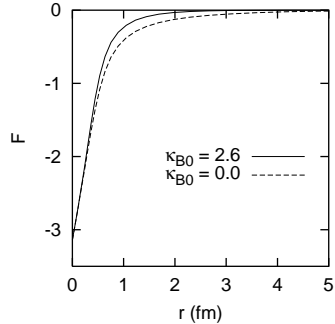}
		\caption{The radial component of the profile function $F$ with $\abs{\kappa_{B3}} = 0$.
		The solid line represents the function at $\kappa_{B0} = 2.6$
		and the dashed line at $\kappa_{B0} = 0$.}
\label{fig:profile1}
	\end{center}
\end{figure}

\begin{figure}
	\begin{center}
		\includegraphics{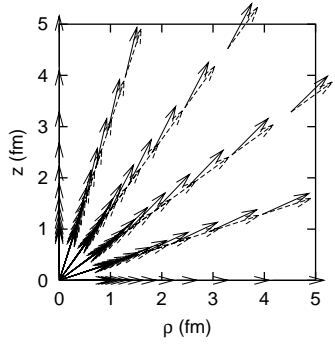}
		\caption{The unit vector $\hat{\bf \Lambda}$ in isospin space
		with $\abs{\kappa_{B3}} = \kappa_{B0}$ and $\rho = \sqrt{x^{2}+y^{2}}$.
		The solid arrow represents the vector at $\kappa_{B0} = 2.6$
		and the dashed arrow at $\kappa_{B0} = 0$.}
\label{fig:profile2}
	\end{center}
\end{figure}

Both the profile functions and the classical soliton energy
are some even functions of $\kappa_{B3}$ because of the isospin symmetry of the model.
The value of $\kappa_{B0}$ mainly affects the $r$ dependence of the profile functions
and the value of $\kappa_{B3}$ affects the $\theta$ dependence of the profile functions.
It was found from the calculation that $F$ hardly depends on $\abs{\kappa_{B3}}$.
Thus, we can treat $F$ as a function of $r$ only.
The profile $F({\bf r})$ is shown for the cases of $\kappa_{B0} = 0, 2.6$
with $\abs{\kappa_{B3}} = 0$ in Fig.~\ref{fig:profile1}.
It is shown below that the range $\kappa_{B0} = 0-2.6$ corresponds
to the octet baryon.
On the other hand, the unit vector $\hat{\bf \Lambda}$  has
axial symmetry about the $z$ axis and plane symmetry with respect to the $xy$ plane.
The deviation from the hedgehog shape is an increasing function of $\abs{\kappa_{B3}}$
and takes a maximum at $\theta = \pi/4,3\pi/4$ 
and a minimum at $\theta = 0,\pi/2,\pi$ in real space.
We show $\hat{\bf \Lambda}$ at $\kappa_{B0} = \kappa_{B3} = 0, 2.6$
in Fig.~\ref{fig:profile2}.
\begin{figure}
	\begin{center}
		\includegraphics{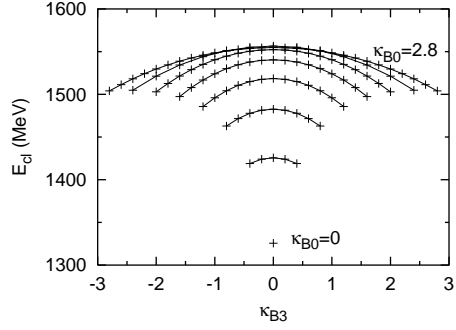}
		\caption{Classical soliton energy.
		Each curve corresponds to a value of $\kappa_{B0} = 0-2.8$.
		The range of $\kappa_{B3}$ is restricted to
		$- \kappa_{B0} \leq \kappa_{B3} \leq + \kappa_{B0}$.}
\label{fig:ecl_2d}
	\end{center}
\end{figure}

Figure~\ref{fig:ecl_2d} displays the $\kappa_{B3}$ dependence of $E_{cl}$.
Each curve corresponds to a value of $\kappa_{B0}$, and the range of $\kappa_{B3}$
is restricted by Eq.~(\ref{eq:ddTrange}).
The classical soliton energy $E_{cl}$ is an increasing function of $\kappa_{B0}$
around $\kappa_{B0} = 0$ and a decreasing function of $\abs{\kappa_{B3}}$ at a fixed $\kappa_{B0}$.
Because the deviation from the hedgehog shape increases with $\abs{\kappa_{B3}}$,
the soliton takes a stable deformed shape at $\kappa_{B0} \neq 0$.

\subsection{Effective Lagrangian and Hamiltonian}
There are two expansion parameters for the effective action $S_{F}$:
the number of colors $N_{c}$ and the SU(3) symmetry breaking $m_{8}$. 
We expand $S_{F}$ up to second order in powers of $1/N_{c}$ 
and first order in powers of $m_{8}$.

First, we define the local variables $\alpha$, $a$, and $b$
of the rotation ${\cal A}$, $A$, and $B$, respectively, and the fluctuation $\sigma$ by
\begin{eqnarray}
	\dot{\alpha}^{\mu} T_{\mu} &=& -i {\cal A}^{\dagger}\dot{\cal A}\nonumber\\
		&=& A_{s}^{\dagger} \left(\begin{array}{cc}
					- i A^{\dagger}\dot{A} & 0\\
					0^{\dagger} & 0
			\end{array}\right) A_{s} - i A_{s}^{\dagger}\dot{A_{s}},\\
	\dot{a}^{j} \frac{\tau_{j}}{2} &=& -i A^{\dagger} \dot{A},\\
	\dot{b}^{j} J^{(q)}_{j} &=& -i B \dot{B}^{\dagger},\\
	\sigma^{\mu} T_{\mu} &=& \hat{m}' - \hat{m}_{B},
\end{eqnarray}
where $T_{\mu} = \lambda_{\mu}/2$, $\mu = 1,2,\dots,8$, and $j = 1,2,3$.
Since in the large $N_{c}$ limit \cite{rf:tHooft74,rf:Witten79},
\begin{eqnarray}
	E_{cl} &\sim& N_{c},\\
	i A^{\dagger} \dot{A} &\sim& 1/N_{c},\\
	i B \dot{B}^{\dagger} &\sim& 1/N_{c},\\
	D,\,D^{\dagger} &\sim& 1/\sqrt{N_{c}},
\end{eqnarray}
we expand $S_{F}$ up to second order in $\dot{\alpha}$ and $\dot{b}$,
and first order in $\sigma$.
Thus, we get the Lagrangian \cite{rf:Reinhardt89,rf:Weigel92}
\begin{eqnarray}
	{\cal L}= \frac{S_{F}}{T} &=& - E_{cl}
		+ \frac{1}{2} \dot{\alpha}^{\mu} \dot{\alpha}^{\nu} U_{\mu\nu}
			+ \frac{1}{2} \dot{b}^{i} \dot{b}^{j} V_{ij} 
			- \dot{\alpha}^{\mu} \dot{b}^{j} W_{{\mu}j}\nonumber\\
		&&+ \dot{\alpha}^{\mu} \sigma^{\nu} \Delta_{\mu\nu}
			- \dot{b}^{i} \sigma^{\nu} \tilde{\Delta}_{i{\mu}} \nonumber\\
		&&- \dot{\alpha}^{\mu} B_{\mu} + \dot{b}^{i} \tilde{B}_{i}
			- \sigma^{\mu} \Gamma_{\mu},
\label{eq:leff_g}
\end{eqnarray}
where the coefficients are defined as
\begin{eqnarray}
	&&U_{\mu\nu} = \sum_{m \neq n} \rho_{\Lambda}^{R}(\epsilon_{m},\epsilon_{n})
				\braketo{m}{T_{\mu}}{n} \braketo{n}{T_{\nu}}{m},\nonumber\\
	&&V_{ij} = \sum_{m \neq n} \rho_{\Lambda}^{R}(\epsilon_{m},\epsilon_{n})
				\braketo{m}{J^{(q)}_{i}}{n} \braketo{n}{J^{(q)}_{j}}{m},\nonumber\\
	&&W_{{\mu}j} = - \sum_{m \neq n} \rho_{\Lambda}^{R}(\epsilon_{m},\epsilon_{n})
				\braketo{m}{T_{\mu}}{n} \braketo{n}{J^{(q)}_{j}}{m},\nonumber\\
	&&B_{\mu} = \sum_{m} \rho_{\Lambda}^{I}(\epsilon_{m}) \braketo{m}{T_{\mu}}{m},\nonumber\\
	&&\tilde{B}_{i} = - \sum_{m} \rho_{\Lambda}^{I}(\epsilon_{m}) \braketo{m}{J^{(q)}_{i}}{m},\nonumber\\
	&&\Delta_{\mu\nu} = \sum_{m \neq n} \rho_{\Lambda}^{I}(\epsilon_{m},\epsilon_{n})
				\braketo{m}{T_{\mu}}{n} \braketo{n}{\beta T_{\nu}}{m},\nonumber\\
	&&\tilde{\Delta}_{i{\nu}} = - \sum_{m \neq n} \rho_{\Lambda}^{I}(\epsilon_{m},\epsilon_{n})
				\braketo{m}{J^{(q)}_{i}}{n} \braketo{n}{\beta T_{\nu}}{m},\nonumber\\
	&&\Gamma_{\mu} = \sum_{m} \rho_{\Lambda}^{R}(\epsilon_{m})
				\braketo{m}{\beta T_{\mu}}{m}.
\end{eqnarray}
Here, $\rho_{\Lambda}^{R,I}(\epsilon_{m})$ and $\rho_{\Lambda}^{R,I}(\epsilon_{m},\epsilon_{n})$ are
the cutoff functions shown in the Appendix.
The indices $R$ and $I$ denote the origin of the vacuum part of the coefficients from the real and
imaginary parts of the action $S_{F}$ in the imaginary time prescription.
We subtract the vacuum contribution of the eigenstates of $K'$[Eq.~(\ref{eq:kinetic})]
from these quantities in a similar way to $E_{cl}$.

By $K_{3}$ symmetry, Eq.~(\ref{eq:leff_g}) becomes
\begin{widetext}
\begin{eqnarray}
	{\cal L} &=& - E_{cl}
		+ \frac{1}{2} \left(
			\dot{\alpha}^{1}\dot{\alpha}^{1} 
			+ \dot{\alpha}^{2}\dot{\alpha}^{2}
		\right) U_{11}
		+ \frac{1}{2} \left(\dot{b}^{1}\dot{b}^{1}
		+ \dot{b}^{2}\dot{b}^{2}\right) V_{11}
		- \left(\dot{\alpha}^{1} \dot{b}^{1} 
		+ \dot{\alpha}^{2} \dot{b}^{2}\right) W_{11}\nonumber\\
		&&+ \left(
			\dot{\alpha}^{1} \sigma^{1} + 
			\dot{\alpha}^{2} \sigma^{2} 
		\right) \Delta_{11}
		- \left(
			\dot{b}^{1} \sigma^{1} +
			\dot{b}^{2} \sigma^{2} 
		\right) \tilde{\Delta}_{11}\nonumber\\
		&&+ \frac{1}{2} \left(\dot{\alpha}^{3} - \dot{b}^{3}\right)^2 U_{33}
		- \left(\dot{\alpha}^{3} - \dot{b}^{3}\right) 
			\left(B_{3} - \sigma^{3} \Delta_{33} - \sigma^{8} \Delta_{38}\right)
		- \dot{\alpha}^{8} B_{8} - \sigma^{3} \Gamma_{3} - \sigma^{8} \Gamma_{8}\nonumber\\
		&&+ \frac{1}{2} \left(
			\dot{\alpha}^{4}\dot{\alpha}^{4} 
			+ \dot{\alpha}^{5}\dot{\alpha}^{5}
		\right) U_{44}
		+ \frac{1}{2} \left(
			\dot{\alpha}^{6}\dot{\alpha}^{6} 
			+ \dot{\alpha}^{7}\dot{\alpha}^{7}
		\right) U_{66}
		+ \left(
			\dot{\alpha}^{4} \sigma^{4} + 
			\dot{\alpha}^{5} \sigma^{5} 
		\right) \Delta_{44}
		+ \left(
			\dot{\alpha}^{6} \sigma^{6} + 
			\dot{\alpha}^{7} \sigma^{7} 
		\right) \Delta_{66}.
\label{eq:leff}
\end{eqnarray}
\end{widetext}
Since $\dot{\alpha}$ and $\sigma$ in Eq.~(\ref{eq:leff_g}) are
functions of $D$,\,$D^{\dagger}$ and $i A^{\dagger} \dot{A}$,
we should further expand these quantities in the actual calculation.

Next, we would like to express the Hamiltonian in terms of $a$,$b$,$D$ and
their canonically conjugate momenta.
The canonical momenta conjugate to these coordinates are defined by 
\begin{eqnarray}
	I_{i} &=& \frac{\partial {\cal L}}{\partial \dot{a}^{i}},
\label{eq:I_coll}\\
	J_{i} &=& \frac{\partial {\cal L}}{\partial \dot{b}^{i}},
\label{eq:J_coll}\\
	P^{\gamma} &=& \frac{\partial {\cal L}}{\partial \dot{D}_{\gamma}^{\dagger}}
	\,\,\,(\gamma = 1, 2),
\label{eq:P_coll}
\end{eqnarray}
and satisfy the following commutation relations:
\begin{eqnarray}
	\comm{I_{i}}{a^{j}} &=& \comm{J_{i}}{b^{j}}
		= \frac{1}{i} \delta_{i}^{j},\\
	\comm{P^{\gamma}}{D^{\dagger}_{\beta}}
	&=& \comm{P^{\dagger}_{\beta}}{D^{\gamma}}
		= \frac{1}{i} \delta^{\gamma}_{\beta}.
\end{eqnarray}

The collective Hamiltonian derived from Eq.~(\ref{eq:leff}),
\begin{equation}
	{\cal H} = \dot{\bm a} \cdot {\bf I} + \dot{\bf b} \cdot {\bf J}
		+ P^{\dagger}\dot{D} + \dot{D}^{\dagger} P - {\cal L},
\label{eq:heff}
\end{equation}
is calculated up to order $1/N_{c}$.
Because of the $K_{3}$ symmetry of the chiral fields,
there is the following constraint on the canonical momenta:
\begin{equation}
	I_{3} + J_{3} + I_{K3} = 0.
\label{eq:constraint_3}
\end{equation}
Here $I_{K3}$ is the third component of the isospin carried by the kaon, 
\begin{equation}
	{\bf I}_{K} = i \left( D^{\dagger} \frac{{\bm \tau}}{2} P 
			-P^{\dagger} \frac{{\bm \tau}}{2} D \right).
\label{eq:isospin_s}
\end{equation}
If we assume the hedgehog shape for the chiral fields,
the constraint becomes
\begin{equation}
	{\bf I} + {\bf J} + {\bf I}_{K} = 0.
\label{eq:constraint}
\end{equation}

\section{Diagonalization of the collective Hamiltonian\label{sec:diagonal}}
\subsection{The mean field approximation\label{sec:kaon}}
Before diagonalizing $\cal H$ [Eq.~(\ref{eq:heff})],
we will describe ${\cal H}_{0}$, which is the $O(1)$ Hamiltonian of $\cal H$
in the large $N_{c}$ limit in the mean field approximation:
\begin{widetext}
\begin{eqnarray}
	{\cal H}_{0} &=& E_{cl} + E_{ind}
		+\frac{1}{4 \Phi_{(+)}} P^{\dagger} P
		+\frac{1}{4 \Phi_{(-)}} P^{\dagger} \tau_{3} P\nonumber\\
		&&+ 3 \left[
			\frac{B_{8}^{2} }{4 \Phi_{(+)}} 
			+2 m_{8} \left(\Delta_{(+)} B_{8}-\Gamma_{8}\right)
		\right] D^{\dagger} D
		+ \sqrt{3} \left(\frac{B_{8}}{4 \Phi_{(+)}} + m_{8} \Delta_{(+)}\right)
			i \left(D^{\dagger}P-P^{\dagger}D\right) \nonumber\\
		&&+ 3 \left[
			\frac{B_{8}^{2}}{4 \Phi_{(-)}} 
			+2 m_{8} \left(\Delta_{(-)} B_{8}-\frac{\Gamma_{3}}{\sqrt{3}}\right)
		\right] D^{\dagger} \tau_{3} D
		+ \sqrt{3} \left(\frac{B_{8}}{4 \Phi_{(-)}} + m_{8} \Delta_{(-)}\right)
			i \left(D^{\dagger}\tau_{3}P-P^{\dagger}\tau_{3}D\right),\nonumber\\
\label{eq:h_N0}
\end{eqnarray}
\end{widetext}
where
\begin{eqnarray}
	E_{ind} &=& 3 m_{8} \left(
			\kappa_{B0} \Gamma_{8} + \kappa_{B3} \frac{\Gamma_{3}}{\sqrt{3}}
	\right) \frac{\sin^2 \sqrt{\kappa_{B0}}}{\kappa_{B0}},\\
	\frac{1}{\Phi_{(\pm)}} &=& \frac{1}{2}
		\left(\frac{1}{U_{44}} \pm \frac{1}{U_{66}}\right),\\
	\Delta_{(\pm)} &=& \frac{1}{2}
		\left(\frac{\Delta_{44}}{U_{44}} \pm \frac{\Delta_{66}}{U_{66}}\right).
\end{eqnarray}
The eigenstate of ${\cal H}_{0}$ describes the kaon in the background soliton.
The energy $E_{ind}$ is induced by the deviation to the strange 
and isospin direction from the SU(2) hedgehog soliton.
Here the $\Gamma_{8}$ term in $E_{ind}$ gives a negative contribution to ${\cal H}_{0}$ and 
$\Gamma_{3}$ a spositive one.

The Hamiltonian ${\cal H}_{0}$ is in a bilinear form of the isodoublets $D$ and $P$,
and the type of separation of variables with respect to the individual component.
For the individual components, the Hamiltonian except for the constant terms is given by
\begin{eqnarray}
	{\cal H}_{\gamma} &=& h_{1\gamma} P^{\dagger}_{\gamma} P^{\gamma} 
		  + h_{2\gamma} D^{\dagger}_{\gamma} D^{\gamma}\nonumber\\
		&&+ h_{3\gamma} (P^{\dagger}_{\gamma} D^{\gamma} + D^{\dagger}_{\gamma} P^{\gamma})\nonumber\\
		&&+ h_{4\gamma} i (P^{\dagger}_{\gamma} D^{\gamma} - D^{\dagger}_{\gamma} P^{\gamma}),
		\,\,\, (\gamma = 1,2)
\end{eqnarray}
where $h_{i\gamma}$ are constants for each component.
If $h_{i\gamma}$ satisfy
\begin{equation}
	h_{1\gamma} h_{2\gamma} - (h_{3\gamma})^{2} \geq 0,
\label{eq:hcoff_cond}
\end{equation}
${\cal H}_{0}$ can be diagonalized exactly using the following transformations \cite{rf:Westerberg94}:
\begin{eqnarray}
	D^{\gamma} &=& c_{\gamma} \left( \xi_{\gamma} + \bar{\xi}^{\dagger}_{\gamma}\right),\\
	P^{\gamma} &=& \frac{1}{2i c_{\gamma} \cos \delta_{\gamma}}
		\left( e^{-i \delta_{\gamma}} \xi_{\gamma} 
			- e^{+i \delta_{\gamma}} \bar{\xi}^{\dagger}_{\gamma}\right),
\end{eqnarray}
where $c_{\gamma}$ and $\delta_{\gamma}$ are constants depending on $h_{i\gamma}$.
The quantities $\xi_{\gamma}^{\dagger}$ ($\xi_{\gamma}$)
and $\bar{\xi}_{\gamma}^{\dagger}$ ($\bar{\xi}_{\gamma}$) are
the creation (annihilation) operators for states with the same quantum numbers
as the kaon and antikaon, respectively.
These satisfy the commutation relations
\begin{eqnarray}
	\comm{\xi_{\gamma}}{\xi_{\beta}^{\dagger}} &=&
	\comm{\bar{\xi}_{\gamma}}{\bar{\xi}_{\beta}^{\dagger}} = \delta_{\gamma\beta},\\
	\comm{\xi_{\gamma}}{\bar{\xi}_{\beta}} &=&
	\comm{\xi_{\gamma}}{\bar{\xi}_{\beta}^{\dagger}} = 0.
\end{eqnarray}

This then finally leads to the diagonalized form
\begin{eqnarray}
	{\cal H}_{0} &=& E_{cl} + E_{ind}
	+ \sum_{\gamma = 1}^{2} \left(
		\omega_{\gamma} \xi_{\gamma}^{\dagger} \xi_{\gamma}
		+ {\bar \omega}_{\gamma}  \bar{\xi}_{\gamma}^{\dagger} \bar{\xi}_{\gamma}
	\right),
\label{eq:h_N0_diag}
\end{eqnarray}
where
\begin{eqnarray}
	\omega_{\gamma} &=& {\rm sgn}(h_{1\gamma})
		\sqrt{h_{1\gamma} h_{2\gamma} - (h_{3\gamma})^{2}} - h_{4\gamma},\\
	{\bar \omega}_{\gamma} &=& {\rm sgn}(h_{1\gamma})
		\sqrt{h_{1\gamma} h_{2\gamma} - (h_{3\gamma})^{2}} + h_{4\gamma}.
\end{eqnarray}
There are also two physical quantities diagonalized
by the creation and annihilation operators:
the third component of the isospin carried by the kaon [Eq.~(\ref{eq:isospin_s})],
\begin{equation}
	I_{K3} = \frac{1}{2} \left(
		\xi_{1}^{\dagger} \xi_{1} - \bar{\xi}_{1}^{\dagger} \bar{\xi}_{1}
		- \xi_{2}^{\dagger} \xi_{2} + \bar{\xi}_{2}^{\dagger} \bar{\xi}_{2}
	\right),
\label{eq:isospin}
\end{equation}
and the strangeness carried by the kaon,
\begin{equation}
	S = i \left(D^{\dagger} P -P^{\dagger} D\right)
		= \sum_{\gamma = 1}^{2} \left(
			\xi_{\gamma}^{\dagger} \xi_{\gamma}
			- \bar{\xi}_{\gamma}^{\dagger} \bar{\xi}_{\gamma}
		\right).
\label{eq:strangeness}
\end{equation}
The Fock space is generated by successive operation of the creation operators
on the vacuum state $\ket{0}$:
\begin{equation}
	\ket{n_{1}, \bar{n}_{1}, n_{2}, \bar{n}_{2}} 
	= \prod_{\gamma = 1}^{2} \frac{1}{\sqrt{n_{\gamma}!\, \bar{n}_{\gamma}!}}
	\left(\xi_{\gamma}^{\dagger}\right)^{n_{\gamma}}
	\left(\bar{\xi}_{\gamma}^{\dagger}\right)^{\bar{n}_{\gamma}} \ket{0},
\label{eq:fock}
\end{equation}
where $n_{\gamma}$ and $\bar{n}_{\gamma}$ are some positive integers.

In the mean field approximation, $\kappa_{B0}$ and $\kappa_{B3}$ should be
self-consistently determined by Eqs.~(\ref{eq:ddT0}) and (\ref{eq:ddT3}).
Then the stability of the approximation should be checked.
Because the coefficients of the individual terms of ${\cal H}_{0}$ are evaluated using 
a soliton depending on $\kappa_{B0}$ and $\kappa_{B3}$, the potential term is 
physically meaningful only in the vicinity of the expectation values.
It is difficult to draw a potential diagram over a wide range of $D$ and $D^{\dagger}$.
The reason is that ${\cal H}_{0}$ describes the combined system 
of the classical soliton and the kaon. 

\begin{figure}
	\includegraphics{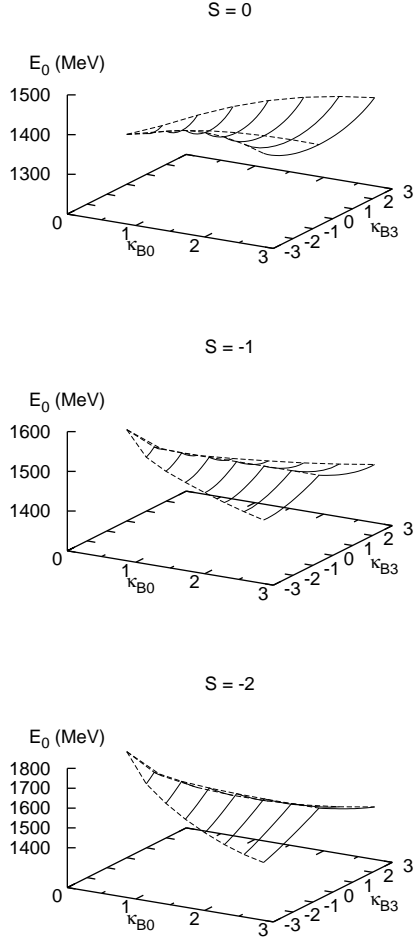}
	\caption{The lowest eigenvalues of ${\cal H}_{0}$, where $\kappa_{B0}$
	and $\kappa_{B3}$ are treated as parameters.}
\label{fig:mean_field}
\end{figure}

In order to investigate the stability of the mean field approximation
we will treat $\kappa_{B0}$ and $\kappa_{B3}$ as parameters at first.
When $\kappa_{B0}$ and $\kappa_{B3}$ are different from the expectation values
of Eqs.~(\ref{eq:ddT0}) and (\ref{eq:ddT3}), the differences
\begin{eqnarray}
 	\kappa_{B0} &-& \braketo{B}{\kappa_{0}}{B},\\
	\kappa_{B3} &-& \braketo{B}{\kappa_{3}}{B}
\end{eqnarray}
act on the system as a kind of external field.
Therefore, by investigating the $\kappa_{B0}$ and $\kappa_{B3}$ dependencies
of the lowest eigenvalue $E_{0}$ of ${\cal H}_{0}$,
we can study the stability of the system against external perturbation.
Figure~\ref{fig:mean_field} shows the behavior of $E_{0}$ as a function of
$\kappa_{B0}$ and $\kappa_{B3}$ in the cases of $S = 0, -1, -2$.
$E_{0}$ is an even function of $\kappa_{B3}$ like $E_{cl}$.

\begin{figure}
	\begin{center}
		\includegraphics{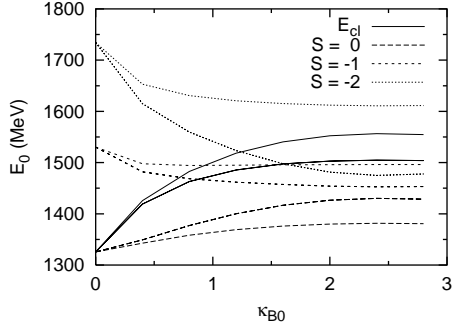}
	\end{center}
	\caption{The lowest eigenvalues of ${\cal H}_{0}$ and the classical soliton energy
	in the cases of $\abs{\kappa_{B3}} = 0$ and $\abs{\kappa_{B3}} = \kappa_{B0}$. 
	Between the two curves with $S = 0$, the case of $\abs{\kappa_{B3}} = 0$
	corresponds to lower energy. For $S = -1, -2$ it corresponds to higher energy.}
\label{fig:e_N0_dd0}
\end{figure}

We further investigate the characteristic behavior of $E_{0}$ in the three cases.
Figure~\ref{fig:e_N0_dd0} shows the $\kappa_{B0}$ dependence of $E_{0}$.
For $S = 0$, the $\kappa_{B0}$ dependence of $E_{0}$ is similar to that of 
the SU(2) sector quark mass $3 (m_{Bu}+m_{Bd})/2$.
It is due to the $\Gamma_{8}$ term in $E_{ind}$ of Eq.~(\ref{eq:h_N0})
that $E_{0}$ is flat compared with $E_{cl}$ or $3 (m_{Bu}+m_{Bd})/2$.
In the case of $S = -1$, $E_{0}$ shows a similar tendency to $m_{Bu} + m_{Bd} + m_{Bs}$.
In the case of $S = -2$, $E_{0}$ is similar to $(m_{Bu}+m_{Bd})/2 + 2 m_{Bs}$. 

\begin{figure}
	\begin{center}
		\includegraphics{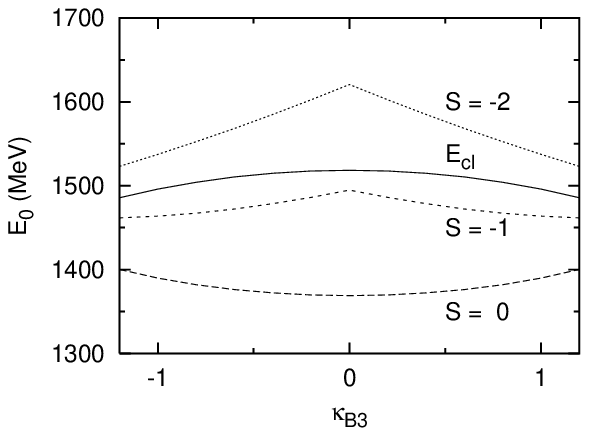}
	\end{center}
	\caption{The lowest eigenvalues of ${\cal H}_{0}$ and the classical soliton energy
	at $\kappa_{B0} = 1.0$.}
\label{fig:e_N0_dd3}
\end{figure}

Figure~\ref{fig:e_N0_dd3} shows the $\kappa_{B3}$ dependence of $E_{0}$.
It is due to $\Gamma_{3}$ term in $E_{ind}$
that $E_{0}$ is convex downward as a function of $\kappa_{B3}$ for $S = 0$.
For $S = -1, -2$, since $E_{ind}$ is canceled by the potential of $D$ and $D^{\dagger}$,
$E_{0}$ is convex upward due to the $\kappa_{B3}$ dependence of $m_{Bu,d}$ (Fig.~\ref{fig:meff}).

Therefore, in the case of $S = 0$ it is energetically forbidden
that the soliton deviate from the hedgehog shape,
and the harmonic analysis around $\kappa_{B0} = \kappa_{B3} = 0$ is justified.
On the other hand, in cases of $S = -1, -2$ $E_{0}$ is energetically
unstable at $\kappa_{B3} = 0$. Thus the soliton deviates from the hedgehog shape.

However, it is necessary to keep in mind that 
under the condition of Eq.~(\ref{eq:hcoff_cond})
the potential term of ${\cal H}_{0}$ is convex downward and
instability for $D$,\,$D^{\dagger}$ is not caused.
Thus, the instability is due to the classical soliton.
If the degrees of freedom for the $\kappa_{B3}$ direction are not included
in the soliton, the instability does not occur.

In Table~\ref{tb:mean_field} we show $\kappa_{B0}$, {\abs{\kappa_{B3}}} and
$E_{0}(\equiv E_{mf})$, which are self-consistently determined in the mean field approximation.
\begin{table}
\caption{$\kappa_{B0}$, \abs{\kappa_{B3}}, and $E_{mf}$ for $S = 0, -1, -2$
	in the mean field approximation.}
	\begin{ruledtabular}
	\begin{tabular}{ccccc}
	$S$ & $\kappa_{B0}$ & $|\kappa_{B3}|$ & $E_{cl}$~(MeV) & $E_{mf}$~(MeV)\\
	\hline
	0	& 0	&   0& 1326 & 1326\\
	-1	& 0.8	& 0.8& 1463 & 1468\\
	-2	& 2.6	& 2.6& 1505 & 1475\\
	\end{tabular}
	\end{ruledtabular}
\label{tb:mean_field}
\end{table}

\subsection{Baryon as the rotational band}
The Hamiltonian ${\cal H}$ [Eq.~(\ref{eq:heff})] is a highly complicated function
of $U$,$V$, etc.
Thus we focus on the algebraic point of view
and do not show its exact form at this point.
The actual calculation will be performed numerically.

For the diagonalization of ${\cal H}$,
we introduce the eigenstates of the body fixed operators
$\hat{{\bf J}}^{2}$,$\hat{J}_{3}$,$\hat{{\bf I}}^{2}$,$\hat{I}_{3}$
defined by Eqs.~(\ref{eq:I_coll}) and (\ref{eq:J_coll}),
and the space fixed spin $\hat{j}_{3}$ and isospin $\hat{i}_{3}$ operators.
Hereafter we denote the operators by a character with a caret 
and a character without a caret denotes its eigenvalue:
\begin{eqnarray}
	\hat{{\bf J}}^{2} \ket{J, j_{3}, J_{3};\, I, i_{3}, I_{3}}
		&=& J(J+1) \ket{J, j_{3}, J_{3};\, I, i_{3}, I_{3}},\nonumber\\
	\hat{j}_{3} \ket{J, j_{3}, J_{3};\, I, i_{3}, I_{3}}
		&=& j_{3} \ket{J, j_{3}, J_{3};\, I, i_{3}, I_{3}},\nonumber\\
	\hat{J}_{3} \ket{J, j_{3}, J_{3};\, I, i_{3}, I_{3}}
		&=& J_{3} \ket{J, j_{3}, J_{3};\, I, i_{3}, I_{3}},\nonumber\\
	\hat{{\bf I}}^{2} \ket{J, j_{3}, J_{3};\, I, i_{3}, I_{3}}
		&=& I(I+1) \ket{J, j_{3}, J_{3};\, I, i_{3}, I_{3}},\nonumber\\
	\hat{i}_{3} \ket{J, j_{3}, J_{3};\, I, i_{3}, I_{3}}
		&=& i_{3} \ket{J, j_{3}, J_{3};\, I, i_{3}, I_{3}},\nonumber\\
	\hat{I}_{3} \ket{J, j_{3}, J_{3};\, I, i_{3}, I_{3}}
		&=& I_{3} \ket{J, j_{3}, J_{3};\, I, i_{3}, I_{3}}\nonumber\\
\end{eqnarray}
with
\begin{eqnarray}
	-J \leq j_{3}, J_{3} \leq J,\\
	-I \leq i_{3}, I_{3} \leq I.
\end{eqnarray}

By using the constraint of Eq.~(\ref{eq:constraint_3})
between $J_{3}$, $I_{3}$, and $I_{K3}$, the basis for the whole space is given by
\begin{equation}
	\ket{J, j_{3}, J_{3};\, I, i_{3}, -(J_{3}+I_{K3})}\, 
	\ket{n_{1}, \bar{n}_{1}, n_{2}, \bar{n}_{2}},
\label{eq:coll_bases}
\end{equation}
where $I_{K3} = (n_{1}-\bar{n}_{1}-n_{2}+\bar{n}_{2})/2$
from Eqs.~(\ref{eq:isospin}) and (\ref{eq:fock}).
Because ${\cal H}$ commutes with $\hat{S}$ [Eq.~(\ref{eq:strangeness})],
the diagonalization is well performed in the subspace with fixed eigenvalues
$(J, j_{3})$, $(I, i_{3})$, $S = n_{1}-\bar{n}_{1}+n_{2}-\bar{n}_{2}$.
For the octet baryon, $J = \abs{j_{3}} = 1/2$.

Corresponding to the expansion of ${\cal H}$ up to order $1/N_{c}$,
we truncate the Fock space of Eq.~(\ref{eq:fock}) on the condition that
\begin{equation}
	n_{1}+\bar{n}_{1}+n_{2}+\bar{n}_{2} \leq \abs{S} + 2.
\end{equation}
Here we introduce the symbols $\ket{K}$ for the states created
by $\xi_{\gamma}^{\dagger}$, and $\ket{\bar{K}}$ for $\bar{\xi}_{\gamma}^{\dagger}$.
The particular states with the same $S$ consist of $\ket{\abs{S} \bar{K}}$ (valence kaon),
$\ket{\abs{S} \bar{K} + K\bar{K}}$ (valence kaon + kaon-antikaon pair), etc.
Since we determined $\kappa_{B0}$ and $\kappa_{B3}$
for the lowest eigenstate $\ket{\abs{S} \bar{K}}$ of ${\cal H}_{0}$ in the mean field approximation,
the state $\ket{\abs{S} \bar{K} + K\bar{K}}$ may be far from the mean field.
Thus, we should not treat equally $\ket{\abs{S} \bar{K}}$ and $\ket{\abs{S} \bar{K} + K\bar{K}}$.
First, we diagonalize ${\cal H}$ in the subspace $\ket{\abs{S} \bar{K}}$.
Next $\ket{\abs{S} \bar{K} + K\bar{K}}$ is treated as a virtual state in the perturbation.

For example, the bases for the $\Sigma$ particle are given by
\begin{eqnarray}
	\left(\begin{array}{l}
		\ket{J, j_{3}, +1/2;\, I, i_{3}, -1} \ket{0, 0, 0, 1} \\
		\ket{J, j_{3}, -1/2;\, I, i_{3},  0} \ket{0, 0, 0, 1} \\
		\ket{J, j_{3}, +1/2;\, I, i_{3},  0} \ket{0, 1, 0, 0} \\
		\ket{J, j_{3}, -1/2;\, I, i_{3}, +1} \ket{0, 1, 0, 0} \\

		\ket{J, j_{3}, +1/2;\, I, i_{3}, -1} \ket{0, 0, 1, 2} \\
		\ket{J, j_{3}, -1/2;\, I, i_{3},  0} \ket{0, 0, 1, 2} \\
		\ket{J, j_{3}, +1/2;\, I, i_{3},  0} \ket{0, 1, 1, 1} \\
		\ket{J, j_{3}, -1/2;\, I, i_{3}, +1} \ket{0, 1, 1, 1} \\
 
		\ket{J, j_{3}, +1/2;\, I, i_{3}, +1} \ket{0, 2, 1, 0} \\
		\ket{J, j_{3}, -1/2;\, I, i_{3}, -1} \ket{1, 0, 0, 2} \\
		\ket{J, j_{3}, +1/2;\, I, i_{3}, -1} \ket{1, 1, 0, 1} \\
		\ket{J, j_{3}, -1/2;\, I, i_{3},  0} \ket{1, 1, 0, 1} \\

		\ket{J, j_{3}, +1/2;\, I, i_{3},  0} \ket{1, 2, 0, 0} \\
		\ket{J, j_{3}, -1/2;\, I, i_{3}, +1} \ket{1, 2, 0, 0} 
	\end{array}\right)
\end{eqnarray}
Here the top four bases span the valence kaon states $\ket{\abs{S} \bar{K}}$,
and the others are bases of $\ket{\abs{S} \bar{K} + K\bar{K}}$.

\begin{figure}
	\begin{center}
	\includegraphics{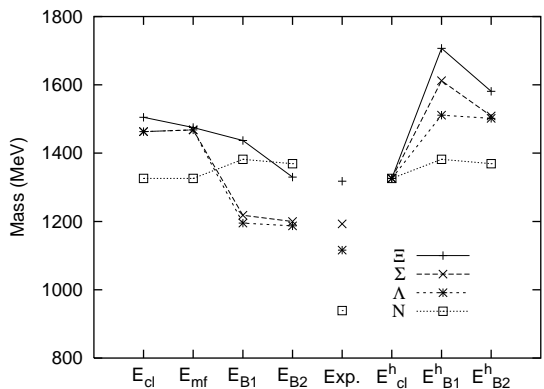}
	\end{center}
	\caption{Comparison of the energy between the deformed soliton 
	and the hedgehog one ($\kappa_{B0}=\kappa_{B3}=0$).
	The superscript $h$ of $E^{h}$ represent the results of the hedgehog soliton. 
	Exp. denotes the experimental values.}
\label{fig:particle}
\end{figure}

Diagonalization of ${\cal H}$ in the basis of Eq.~(\ref{eq:coll_bases})
gives an estimation of the fluctuation for $D$,$J$,$I$
around the mean field approximation.
The results of the calculation are displayed in Table~\ref{tb:particle} 
and compared with the hedgehog results at $\kappa_{B0}=\kappa_{B3}=0$
in Fig.~\ref{fig:particle}.
Here $E_{B1}$ is the eigenvalue of the collective Hamiltonian ${\cal H}$ in only $\ket{\abs{S} \bar{K}}$.
The mass $E_{B2}$ is the eigenvalue in the states that incorporate $\ket{K\bar{K}}$ states as a perturbation.
Except for $N$, the energy eigenvalues $E_{B2}$ are smaller than the hedgehog results
and close to the experimental values.

\begin{table}
\caption{Baryon masses $E_{B1}$ (without $\ket{K\bar{K}}$)
	and $E_{B2}$(with $\ket{K\bar{K}}$).
	Expt. denotes experimental value.}
	\begin{ruledtabular}
	\begin{tabular}{cccc}
	Particle & $E_{B1}$~(MeV)  & $E_{B2}$~(MeV)  & Expt.~(MeV) \\
	\hline
	$N$     	& 1382 & 1369 & 939  \\
	$\Lambda$	& 1195 & 1187 & 1116 \\
	$\Sigma$	& 1218 & 1200 & 1193 \\
	$\Xi$ 		& 1437 & 1330 & 1318 \\
	\end{tabular}
	\end{ruledtabular}
\label{tb:particle}
\end{table}

In the eigenstates of every baryon, there are large transitions among the
bases with different values of $J_{3}+I_{3}$, that is,
remains of the Clebsch-Gordan series for the constraint Eq.~(\ref{eq:constraint})
which is maintained only in the hedgehog shape.
In the case of $N$, the energy eigenvalues are increased by the transition. 
The situation does not change on incorporating the $\ket{K\bar{K}}$ states.
For $\Lambda$ and $\Sigma$, an interaction induced by the deformation of the soliton
picks up the transition. Then the energy shows a decrease.
That is caused only in the space with $\ket{\abs{S}\bar{K}}$.
For $\Xi$, there is a similar decrease in the space
that takes account of $\ket{K\bar{K}}$ perturbatively.
The contributions of $\ket{K\bar{K}}$ to the energy take negative values
for every baryon due to the second order perturbation formula.

\section{Discussion\label{sec:discuss}}
The stability of the hedgehog shape has been investigated
for the octet baryon by means of the chiral quark soliton model.

In the mean field approximation, it was shown that the stable form of 
the soliton changes according to the strangeness of the baryon.
In the case of the soliton without strangeness, the hedgehog shape is stable.
On the other hand, for the soliton with strangeness an instability
occurs not only in the strange direction but also in the isotopic one.
The instability has a global nature due to the inertial force, and
one can not find it from study of the curvature of the local potential.
It is necessary to incorporate the degrees of freedom for the deviation
in the ansatz of the soliton.

After the collective quantization has been performed, 
the states identified as the strange baryons ($\Lambda$,$\Sigma$,$\Xi$)
also energetically favor the deformed soliton.
The strange baryon masses are in good agreement with both the absolute value
and the difference among the experimental values.
These become small compared with the results of the mean field approximation
due to the interaction caused by the deformation of the soliton.
On the other hand, the approach with the hedgehog soliton reproduces 
the mass difference among the baryons well but the absolute value.
Also, the collective states have higher energy than the classical soliton.

Weigel \textit{et al.} investigated the quantum correction due to
the zero modes off the hedgehog soliton in the NJL model \cite{rf:Weigel95,rf:Alkofer95}.
The correction gives a large negative contribution to the $N$ and $\Delta$ masses,c
and the results are in good agreement with the experimental values.
In our approach, the mass of $N$ is somewhat too large due to the hedgehog shape of the soliton.
Therefore, both approaches cooperatively would be in agreement
with the experimental value for $N$.
On the other hand, the deformed soliton corresponding to the strange baryon
has less symmetry than the hedgehog soliton.
Because of the estimation \cite{rf:Weigel95,rf:Alkofer95} relied
on the hedgehog shape \cite{rf:Weigel95},
the relation between the two approaches is obscure at present. 

The eigenvalue of the Hamiltonian up to order 1 in the large $N_{c}$ limit
is an even function of $\kappa_{B3}$. Therefore the mean field energy at $\kappa_{B3}$
is degenerate with that at $-\kappa_{B3}$.
A more accurate calculation of the fluctuation off the mean field 
should treat the tunnel effect.

It is in progress to perform the description of the decuplet baryon
and to incorporate the local variation of the kaon wave function.

\begin{acknowledgments}
We would like to thank M.~Kawabata and H.~Kondo
for many fruitful discussions of this problem.
\end{acknowledgments}

\appendix
\section{Cutoff function\label{sec:cutoff}}
The cutoff function of the vacuum energy for the energy level $\epsilon_{n}$ is given by 
\begin{equation}
	\rho^{vac}_{\Lambda}(\epsilon_{n}) =
		N_{c} {\rm sgn}(\epsilon_{n}) \bar{\cal N}_{\Lambda}(\epsilon_{n}) 	
\end{equation}
with the cutoff parameter $\Lambda$, where, using the Schwinger proper time regularization,
$\bar{\cal N}_{\Lambda}$ may be written as
\begin{equation}
	\bar{\cal N}_{\Lambda}(\epsilon_{m}) = \frac{1}{4 \sqrt{\pi}}
	\int^{\infty}_{(\epsilon_{m}/\Lambda)^{2}} \frac{e^{-\rho}}{\rho^{3/2}} d\rho
\end{equation}

The cutoff function for first order matrix element is given by 
\begin{equation}
	\rho_{\Lambda}^{R,I}(\epsilon_{m}) =
		N_{c} \eta^{val}_{m} + N_{c} {\rm sgn}(\epsilon_{m}) {\cal N}_{\Lambda}^{R,I}(\epsilon_{m})
\end{equation}
with
\begin{eqnarray}
	{\cal N}_{\Lambda}^{R}(\epsilon_{m}) &=& -\frac{1}{2} {\rm erfc}(\abs{\epsilon_{m}/\Lambda}),\\
	{\cal N}_{\Lambda}^{I}(\epsilon_{m}) &=& -\frac{1}{2}.
\end{eqnarray}
The indices $R$ and $I$ denote the origin of the function from the real and
imaginary parts of the action $S_{F}$ in the imaginary time prescription.
Similarly, for the second order matrix elements one obtain 
\begin{equation}
	\rho_{\Lambda}^{R,I}(\epsilon_{m},\epsilon_{n}) =
		N_{c} \frac{\eta^{val}_{m}-\eta^{val}_{n}}{\epsilon_{n}-\epsilon_{m}}
		+\frac{N_{c}}{2} f_{\Lambda}^{R,I}(\epsilon_{m},\epsilon_{n})
\end{equation}
where $f_{\Lambda}^{R,I}$ are given by
\begin{eqnarray}
	f_{\Lambda}^{R}(\epsilon_{m},\epsilon_{n}) &=&
		\frac{
			{\rm sgn}(\epsilon_{m}) {\rm erfc}(\abs{\epsilon_{m}/\Lambda})
				-{\rm sgn}(\epsilon_{n}) {\rm erfc}(\abs{\epsilon_{n}/\Lambda})
		}{\epsilon_{m} - \epsilon_{n}}\nonumber\\
		&&-\frac{2\Lambda}{\sqrt{\pi}}
		\frac{e^{(\epsilon_{m}/\Lambda)^{2}}-e^{(\epsilon_{n}/\Lambda)^{2}}}
				{\epsilon_{m}^{2}-\epsilon_{n}^{2}},\\
	f_{\Lambda}^{I}(\epsilon_{m},\epsilon_{n}) &=&
		\frac{{\rm sgn}(\epsilon_{m}) - {\rm sgn}(\epsilon_{n})}{\epsilon_{m} - \epsilon_{n}}.
\end{eqnarray}


\end{document}